# Anomalous Transport in High-Mobility Superconducting SrTiO$_3$ Thin Films

Jin Yue[1,a)], Yilikal Ayino[2], Tristan K. Truttmann[1], Maria N. Gastiasoro[2,3], Eylon Persky[4], Alex Khanukov[4], Dooyong Lee[1], Laxman R. Thoutam[1,5], Beena Kalisky[4], Rafael M. Fernandes[2], Vlad S. Pribiag[2], and Bharat Jalan[1,a)]

[1]Department of Chemical Engineering and Materials Science
University of Minnesota,
Minneapolis, Minnesota 55455, USA

[2]School of Physics and Astronomy
University of Minnesota,
Minneapolis, Minnesota 55455, USA

[3]Current affiliation: ISC-CNR and Department of Physics, Sapienza University of Rome,
Piazzale Aldo Moro 2, 00185, Rome, Italy

[4]Department of Physics and Institute of Nanotechnology and Advanced Materials
Bar-Ilan University,
Ramat Gan 5290002, Israel

[5]Current affiliation: Department of Electronics and Communications Engineering, SR
University, Warangal Urban, Telangana, 506371, India

a) Corresponding author: yuexx129@umn.edu, bjalan@umn.edu

**Abstract**

The study of subtle effects on transport in semiconductors requires high-quality epitaxial structures with low defect density. Using hybrid molecular beam epitaxy (MBE), SrTiO$_3$ films with low-temperature mobility exceeding 42,000 cm$^2$V$^{-1}$s$^{-1}$ at low carrier density of 3 × 10$^{17}$ cm$^{-3}$ were achieved. A sudden and sharp decrease in residual resistivity accompanied by an enhancement in the superconducting transition temperature were observed across the second Lifshitz transition (LT) where the third band becomes occupied, revealing dominant intra-band scattering. These films further revealed an anomalous behavior in the Hall carrier density as a consequence of the antiferrodistortive (AFD) transition and the temperature-dependence of the Hall scattering factor. Using hybrid MBE growth, phenomenological modeling, temperature-dependent transport measurements, and scanning superconducting quantum interference device imaging, we provide critical insights into the important role of inter- vs intra-band scattering and of AFD domain walls on normal-state and superconducting properties of SrTiO$_3$.

Despite over 60 years of research, SrTiO$_3$ (STO) has continued to surprise researchers with new emerging behaviors[1-4]. This is in large part due to the continuous improvement in the materials quality and the ability to make strain-engineered structures with controlled defect densities. Most recent examples include strain-enhanced superconductivity[5,6] due to an interplay between ferroelectricity and superconductivity[7-9], and phonon thermal Hall effect[10].

Bulk cubic STO transforms to a tetragonal structure upon cooling below ~ 110 K[11]. This is accompanied by an out-of-phase rotation of oxygen octahedra referred to as the antiferrodistortive (AFD) transition. Without an external stress/strain field, the rotational axes of these octahedra align themselves along any cartesian axes, leading to three distinct domains. The boundaries between these domains have been shown to be charged, and even ferroelectric based on several characterizations including resonant ultrasonic spectroscopy[12], scanning superconducting quantum interference device (SQUID) imaging[13,14], piezoelectric spectroscopy measurements[2], scanning single-electron transistor (SET) microscopy[15], scanning transmission electron microscopy[7], and low-temperature scanning electron microscopy (LT-SEM) imaging[16]. Domain boundaries have further been reported to modify the metal-insulator transition[17], cause electrical anisotropy[18-20], anomalous piezoelectricity[15], spatial inhomogeneity in superconductivity[21,22] and the softening of phonons[23,24]. Verma et al. reported that a ~ 6 meV transverse optical phonon deformation potential related to the AFD transition is critical for the transport behavior between 10-200 K[25], while Zhou et al. argued that the AFD soft mode has negligible contribution to transport due to the lack of scattering phase space[26].

Recent experimental and theoretical studies have further renewed interest in STO due to the interplay between ferroelectricity and superconductivity. There is a significant ongoing discussion on the presence of a quantum ferroelectric phase transition in the vicinity of the superconducting dome in STO.[3,4,9,27-29] This aspect of the behavior of STO has raised several exciting questions including on the possibility of enhanced transition temperature for superconductivity at domain boundaries, which are known to be ferroelectric as discussed above. Additionally, it also begs the question of which role local defects, such as dislocations, plays for on superconductivity. Dislocations have local stress-fields around them that can potentially induce local ferroelectricity and locally enhance superconductivity[30]. Clearly, progress on these problems requires cleaner samples with improved defect density and controlled extrinsic defects.

Using hybrid molecular beam epitaxy (MBE)[31-33] and through systematic control of doping, we investigate the role of AFD domain boundaries, and multi-band electronic structure on normal and superconducting properties in STO films. We employ phenomenological modeling, temperature-dependent transport measurements, and SQUID imaging to examine their influences on the carrier density and on the superconducting transition temperature, revealing an important role of the Lifshitz transition and of the temperature-dependent Hall scattering factor.

We begin by discussing the structural properties of STO films, which are critical to establishing a credible case for the electronic structure. Figure 1a shows a representative time-dependent reflection high energy electron diffraction (RHEED) intensity oscillations at the onset of growth indicating a layer-by-layer growth mode. Insets show streaky RHEED patterns, and an atomic force microscopy image of the same 60 nm Nd-doped STO/20 nm undoped STO film on STO (001) substrate. There results attest to the atomically smooth film surfaces. An excellent overlap between the film and substrate peaks was observed in the wide-angle X-ray diffraction scan (WAXRD) (figure 1b), yielding an out-of-plane lattice parameter ($a_{OP}$) of 3.905 ± 0.002 Å, identical to that of the bulk single crystal. Films with ~ 2 % La doping yielded the same $a_{OP}$ = 3.905 ± 0.002 Å with a perfect overlap between film and substrates reflections. This observation suggests that the lattice parameter should not be taken as a sensitive measure of point defects in STO. To further emphasize this point, we show in figure 1c the temperature-dependent electron mobility of a Nd-doped STO film with a carrier density of $3 \times 10^{17}$ cm$^{-3}$. For comparison, we also mark on the same plot the highest low-temperature-mobility values exhibited by bulk single crystalline STO[34] and pulsed-laser deposition (PLD)-grown homoepitaxial STO films[35]. It can be seen that despite sharing an identical lattice parameter, our films yielded significantly higher mobility, suggesting improved defect density and lower impurity concentration.

Having established an optimum growth condition, we synthesized a series of 60 nm Nd-doped STO film/20 nm undoped STO/STO (001) by varying the doping density between $10^{19}$ and $10^{20}$ cm$^{-3}$, a range of carrier density where superconductivity is observed in STO. All samples showed metallic behavior between 1.8 K ≤ $T$ ≤ 300 K with no evidence of localization behavior. At $T$ ≤ 100 K, $\rho$ was found to vary as $T^2$ in all samples (figure 2a); the fitting allows us to extract the intercept ($\rho_0$, the residual resistivity) and the coefficient ($A$). The value of $\rho_0$ as a function of room-temperature carrier density ($n_{300K}$) is plotted in figure 2b whereas the corresponding value of $A$ is shown in figure S1. While the physical origin of $T^2$ behavior is still debated[36-38] (figure S1), an intriguing dependence of $\rho_0$ on $n_{300K}$ was observed in figure 2b manifested by a sudden and sharp decrease at a critical density, ~ 3-4 × $10^{19}$ cm$^{-3}$. Interestingly, this value corresponds to the critical density for the second Lifshitz transition[39] in STO where the chemical potential ($\mu_C$) crosses the third of the three electron-like bands of STO. It, therefore, raises an important question of whether the Lifshitz transition plays any role on the normal state transport.

To investigate this question, we calculated the residual resistivity as a function of band filling ($n/n_c$) for a toy two-band model shown in figure 2c (see supplementary information for details). Here, $n_c$ is the critical density for the second Lifshitz transition. Because the band-splitting by tetragonality is small (~ 2 - 5 meV) compared to that caused by spin-orbit (SO) interactions (~ 12 - 30 meV)[40], we treated the lower two bands as a single one (marked by $\alpha = 1$). The third band is labeled by $\alpha = 2$ in the figure 2c. Assuming isotropic intra-band and inter-band scattering, we express the scattering matrix with an intra-band scattering potential $V_{intra}$ and an inter-band potential $V_{inter}$. Figure 2d shows the calculated normalized resistivity ($\rho/\rho_c$) as a function of $n/n_c$ for different values of inter-band to intra-band scattering potential ratio, $V_{inter}/V_{intra}$. Here, $\rho_c$ refers to the residual resistivity at the critical density, $n_c$ corresponding to the Lifshitz transition. A remarkably similar behavior to our experimental data was observed for $V_{inter}/V_{intra} = 0$ revealing a dominant role of intra-band scattering on normal-state transport across the second Lifshitz transition. To shed further light into this finding, we measured the superconducting transition temperature ($T_c$) in these films (figure 2b), which revealed a continuous increase in $T_c$ across the Lifshitz transition. This result is again consistent with the prior theoretical prediction that in the absence of strong inter-band scattering, $T_c$ should increase across Lifshitz transition[41]. Note that this behavior is the opposite of what was seen previously at the first Lifshitz transition for much lower densities in bulk crystals, in which case $T_c$ was found to be suppressed[39]. It is perhaps worth emphasizing that unlike prior work on STO films, this is first systematic report of superconductivity dome in the uniformly-doped STO films as thin as 60 nm, which is attributable to the lower disorder/defects in these films.

We now turn to the discussion of Hall measurements. Figure 3a shows the Hall carrier density obtained from van der Pauw (vdP) measurements ($n_{Hall} = -1/eR_H$, where $R_H$ is the Hall coefficient) as a function of temperature for 60 nm Nd-doped STO films/20 nm undoped STO/STO (001) with different doping densities. The inset shows a linear Hall slope between $- 9\ T \leq B \leq + 9$ T at 300 K and 2.5 K. Regardless of doping density, all samples exhibited an anomalous behavior around 100 K, i.e. with increasing temperature, $n_{Hall}$ first remains unchanged, then increases followed by a decrease, and finally, increases until room-temperature. We also show, in figure 3b, the $n_{Hall}$ at 1.8 K as a function of $n_{Hall}$ at 300 K for all the samples to investigate a potential correlation between carrier densities at room-temperature and low-temperature. Irrespective of doping density, this plot showed a linear relationship passing through origin while revealing ~ 12% of the room-temperature carriers seemingly disappear upon cooling to 1.8 K. This raises several questions. Why the carriers seem to disappear upon cooling? Why is there an anomalous behavior around 100 K? Also, the linear relationship is surprising. What role if any is played by the AFD, which occurs at around 100K?

We first discuss the origin of the anomalous behavior. Conceivably, one may argue that the anomalous behavior is related to the multi-band electronic conduction in STO, which makes the extraction of the carrier density from the Hall data more subtle[42]. We confirmed this is not the case. Even films with single-band occupancy ($n_{Hall}$ = 6.5 × 10$^{17}$ cm$^{-3}$) yielded a similar anomalous behavior (figure S2). Now, we consider the effect of an often overlooked, yet critical - the Hall scattering factor, $r_H$ - on the measured carrier density. $r_H$ is defined as the ratio between the true 3D carrier density ($n_{3D}$) and the experimentally measured Hall carrier density, $n_{Hall}$, i.e. $r_H = \frac{n_{3D}}{n_{Hall}}$, and is directly related to the relaxation time τ through the equation, $r_H = \frac{\langle\tau^2\rangle}{\langle\tau\rangle^2}$ [43,44]. The value of $r_H$ is usually close to 1 and this is perhaps why it has been assumed to be 1 in all the experimental transport studies in STO. In reality, however, the exact value of $r_H$ depends on the band structure and scattering mechanisms. Unfortunately, there is no systematic study of $r_H$ as a function of carrier density and temperature in STO and the only available data in literature is from *ab initi*o calculations, as shown in figure S3[45]. We used a polynomial interpolation to extract 1/$r_H$ as a function of temperature. Since the calculated $r_H$ is not a strong function of carrier density[45] for $n_{3D}$ between $2\times10^{19}$ and $2\times10^{20}$ cm$^{-3}$, we used $r_H$ corresponding to a fixed $n_{3D} = 2\times10^{19}$ cm$^{-3}$ to extract temperature dependent $n_{3D}$ in all our samples. These results are shown in figure S4. Remarkably, the anomalous behavior is not present after accounting for the temperature-dependence of $r_H$, implying that the temperature-dependence of $r_H$ plays a major role in the origin of the anomalous behavior, and also in determining the true carrier density. Given that the anomalous behavior of $n_{Hall}$ and of $r_H$ both occur near the AFD transition temperature of ~100 K, a likely mechanism for both is an AFD-driven change in the relaxation time, for instance due to the appearance of domain walls or changes in the phonon spectrum.

Although the temperature dependence of $r_H$ accounts for the anomalous peak-and-valley pattern exhibited by $n_{Hall}$, a drop in carrier density as a function of decreasing temperature remains present even after correcting for this effect. One possible explanation for this effect is the emergence of AFD-induced charged/polar domain walls, which could localize some of the carriers. To investigate this hypothesis, we performed scanning superconducting quantum-interference device (SQUID) measurements. An AC current is applied through the adjacent corners of the vdP geometry as shown in figure 3c, and the generated magnetic flux is measured by a 0.75 $\mu$m sensing loop (pick-up loop) via a lock-in technique. By mapping the flux near the surface of the sample, a 2D map describing the local current flow distribution is generated. The scanning SQUID result of the sample with $n_{300\,K} = 4.8 \times 10^{19}$ cm$^{-3}$ (the magenta color in figure 3a) at 4.5 K is shown in figure 3d. Some strips and scattered dots were observed, which indicate a modulated current flow. The stripe-like modulations could be attributed to tetragonal domain patterns. Additionally, these stripes did not change after a temperature cycle to 300 K and these stripe modulations persisted to temperatures above 40 K. The low contrast could be due to the high carrier densities in our samples, which provide a high level of screening of potential steps, therefore leading to smaller current-densities contrast[17,46]. As a result, it becomes very likely that not only the anomalous behavior of $n_{Hall}$ (*T*), but also the drop in true density upon cooling are related to the AFD transition. However, why the dependence is linear in figure 3b is still unclear and may be related to an interplay between AFD and Hall scattering factor as a function of carrier density.

To further discuss the impact of AFD on the anomalous $n_{Hall}$ behavior, we measured $n_{Hall}$ of a representative 60 nm Nd-doped STO films/20 nm undoped STO/STO (001) sample as a function of temperature during warming and cooling. As shown in figure 4a, a clear difference between warming and cooling cycles was observed accompanied by the anomalous behavior. For comparison, we also performed the same measurement with the same warming/cooling rate (5 K/min) on a thicker 160 nm Nd-doped STO sample with similar $n_{Hall}$. In contrast, the thicker sample shows the same anomalous behavior but with no measurable differences between the cooling and warming cycles (figure 4b). Given the 3D carrier density is the same in both films, these results suggest that the hysteretic behavior between warming and cooling cycles are not a result of electronic transition, but likely associated with the dynamic process(es) pertaining to the AFD transition resulting in carrier trapping/detrapping at domain boundaries. To probe the dynamic nature of the temperature dependent $n_{Hall}$, we performed Hall scans at finer temperature steps on the same sample for different warming/cooling rates. Since the Hall slope is linear between ± 9 T at all temperatures, a continuous temperature-dependent measurement of $n_{Hall}$ was performed by keeping the field fixed at ± 9 T. Figure 4c shows $n_{Hall}$ as a function of temperature during warming (labeled as 1 and 3) and cooling (labeled as 2 and 4) cycles performed at two different rates, 1 K/min and 5 K/min. Similar to figure 4a, a small difference in $n_{Hall}$ was observed as a function of warming/cooling cycle. Above a critical temperature, *T\**, the temperature-dependent Hall carrier density upon cooling, i.e., [$n_{Hall}$ (*T*)]$_{cooling}$ was identical to that upon warming ($[n_{Hall}(T)]_{warming}$). However, at $T < T^*$, $[n_{Hall}(T)]_{cooling}$ remained consistently higher than $[n_{Hall}(T)]_{warming}$. The difference ($\Delta n_{Hall}$) between $[n_{Hall}(T)]_{cooling}$ and $[n_{Hall}(T)]_{warming}$ is plotted in figure 4d for two different warming/cooling rates. Figure 4d reveals that $n_{Hall}$ strongly depends on the thermal history of the sample. A smaller $T^*$ of ~ 150 K was obtained for slower warming/cooling cycle as opposed to a higher $T^* = $ ~ 175 K for the faster warming/cooling cycle. While the peak in $\Delta n_{Hall}$ occurs at identical temperature ~ 120 K (the expected AFD transition in doped STO)[40], the onset depends on the cooling rate, suggesting the onset of AFD correlations begins at considerably higher temperatures. One possible explanation can be trapping/detrapping of residual oxygen vacancies, which are known to accumulate at the domain walls[47]. Future studies should focus on examining the normal-state transport and superconductivity at domain boundaries which may be substantially different from the global measurements.

In summary, we have investigated the electrical transport properties in hybrid MBE-grown Nd-doped STO films exhibiting low-temperature mobility exceeding 42,000 cm$^2$V$^{-1}$s$^{-1}$ in the low-doped regime where not all bands are occupied. By systematically varying carrier density across the second Lifshitz transition, the important role of intra- over inter-band scattering on the normal-state transport properties and on the superconducting transition temperature was revealed. Moreover, we showed a superconducting dome in uniformly-doped STO films as thin as 60 nm. This study provides an important connection between the AFD transition, structural dynamics, the Hall scattering factor and carrier localization. Future experimental work should focus on a systematic study of local vs. global transport.

**Acknowledgements:**

The authors would like to thank Eric MaCalla for helpful discussion. This work was primarily supported by the U.S. Department of Energy through the University of Minnesota Center for Quantum Materials (CQM) under Grant DE-SC-0016371. MBE growth (J.Y., B.J.), normal-state transport measurements (J.Y., D.L. L.R.T., B.J.), and theoretical modeling (M.N.G., R.M.F.) were supported by the U.S. Department of Energy through the University of Minnesota CQM under Grant DE-SC-0016371. T.K.T. acknowledges partial support from the U.S. Department of Energy through DE-SC002021 and Air Force Office of Scientific Research (AFOSR) through Grant FA9550-21-1-0025. mK transport measurements (Y.A., V.S.P.) were supported by the National Science Foundation (NSF) Materials Research Science and Engineering Center at the University of Minnesota under Award No. DMR-1420013, and the University of Minnesota McKnight Land Grant Professorship. SQUID imaging (E.P., A.K. and B.K.) was supported by the Israeli Science Foundation grant no. ISF-1281/17, and the QuantERA ERA-NET Cofund in Quantum Technologies, Project No. 731473. Parts of this work were carried out at the Minnesota Nano Center, which is supported by the National Science Foundation through the National Nano Coordinated Infrastructure (NNCI) under Award Number ECCS-1542202. Structural characterizations were carried out at the University of Minnesota Characterization Facility, which receives partial support from NSF through the MRSEC program.

**Author contributions:** J.Y. and B.J. conceived the idea and designed the experiments. J.Y. and T.K.T. grew samples and characterized them structurally. J.Y., D.L. and L.R.T. performed normal-state transport measurements. Y.A. performed mK transport measurement under the supervision of V.S.P.. Theoretical modeling was performed by M.N.G. under the guidance of R.M.F.. E.P.,

A.K. and B.K. performed the SQUID imaging. J.Y. and B.J. wrote the manuscript. All authors contributed to the discussion and manuscript preparation.

**Competing interests:** The authors declare no competing interests.

**Data and materials availability:** All data needed to evaluate the conclusions of the paper are present in the paper and/or the Supplementary Materials. Additional data related to this paper may be requested from the authors.

**Figures:**

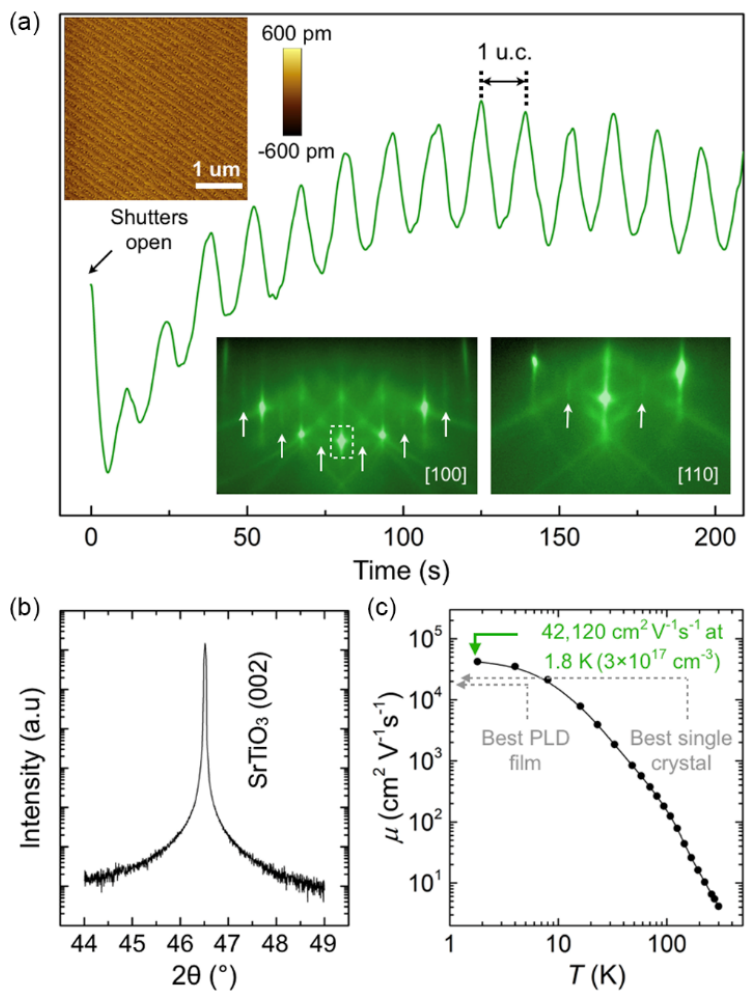

**Figure 1**: (a) Time-dependent RHEED oscillation of a representative 60 nm Nd-doped STO/20 nm undoped STO film on STO (001) substrate. The insets show the AFM image, and the after-growth RHEED patterns along [100] and [110] azimuths. (b) High-resolution X-ray 2 coupled scan of the sample showing an epitaxial, phase pure film. (c) Mobility (μ) vs. temperature plot for a thicker 1060 nm Nd-doped STO/20 nm undoped STO film on STO (001) substrate with carrier density ~ 3 × $10^{17}$ cm$^{-3}$ at 1.8 K with mobility 42,120 cm$^2$V$^{-1}$s$^{-1}$. For comparison, the highest low-temperature mobility value from the single crystal[34] and PLD-grown film[35] is also depicted.

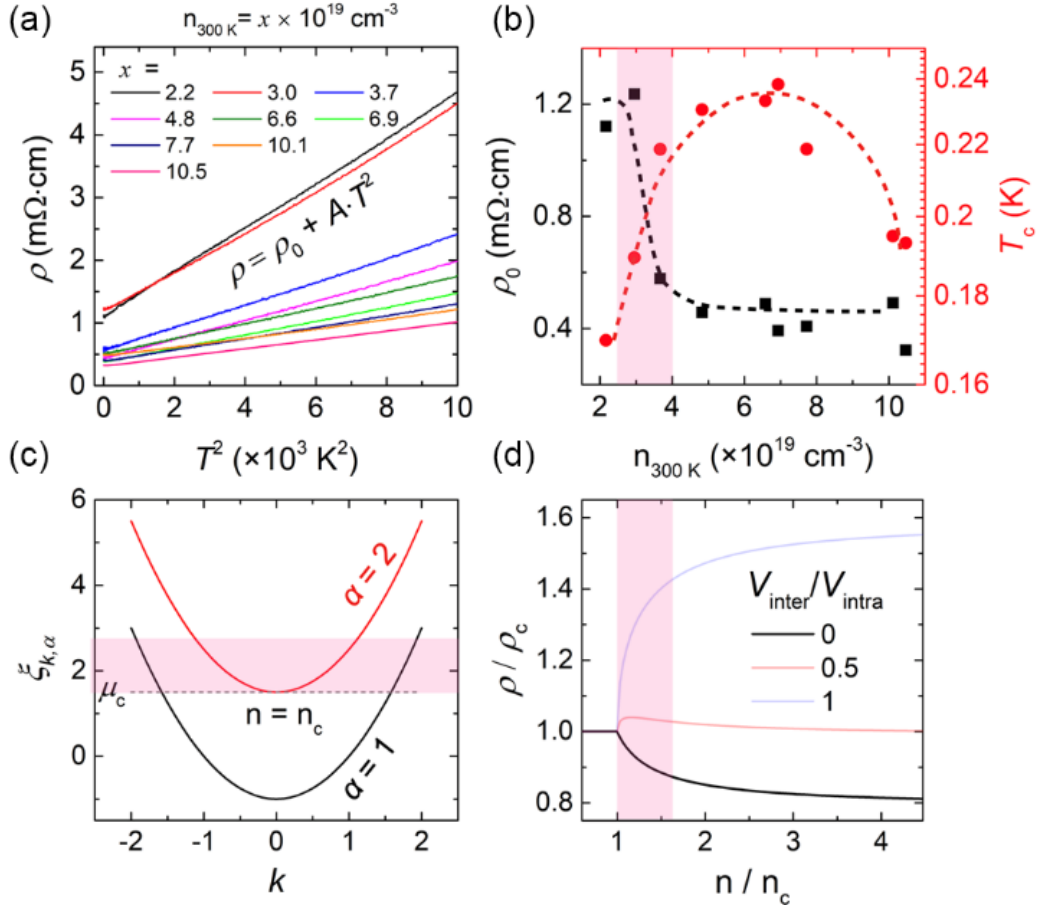

**Figure 2**: (a) Plot of $\rho$ vs $T^2$ plots of 60 nm Nd-doped STO/20 nm undoped STO/STO (001) as a function of carrier density for $T \leq 100$ K. (b) Residual resistivity, $\rho_0$ (left axis) and the superconducting transition temperature, $T_c$ (right axis) as a function of Hall carrier density, $n_{300\ K}$, illustrating a sudden and sharp decrease in $\rho_0$ around $\sim 3 \times 10^{19}$ cm$^{-3}$ (marked by the pink shaded region). Black and red dash lines are guidelines to the eye. (c) Parabolic electronic dispersions of the toy two-band system $\xi_{k,\alpha=1} = k^2/2m - \mu$ and $\xi_{k,\alpha=2} = k^2/2m + \mu_c - \mu$ with the radical momentum, $k$. (d) Calculated normalized residual resistivity $\rho/\rho_c$ vs. normalized carrier density $n/n_c$ for various inter-band to intra-band scattering strength ratios $V_{inter}/V_{intra}$.

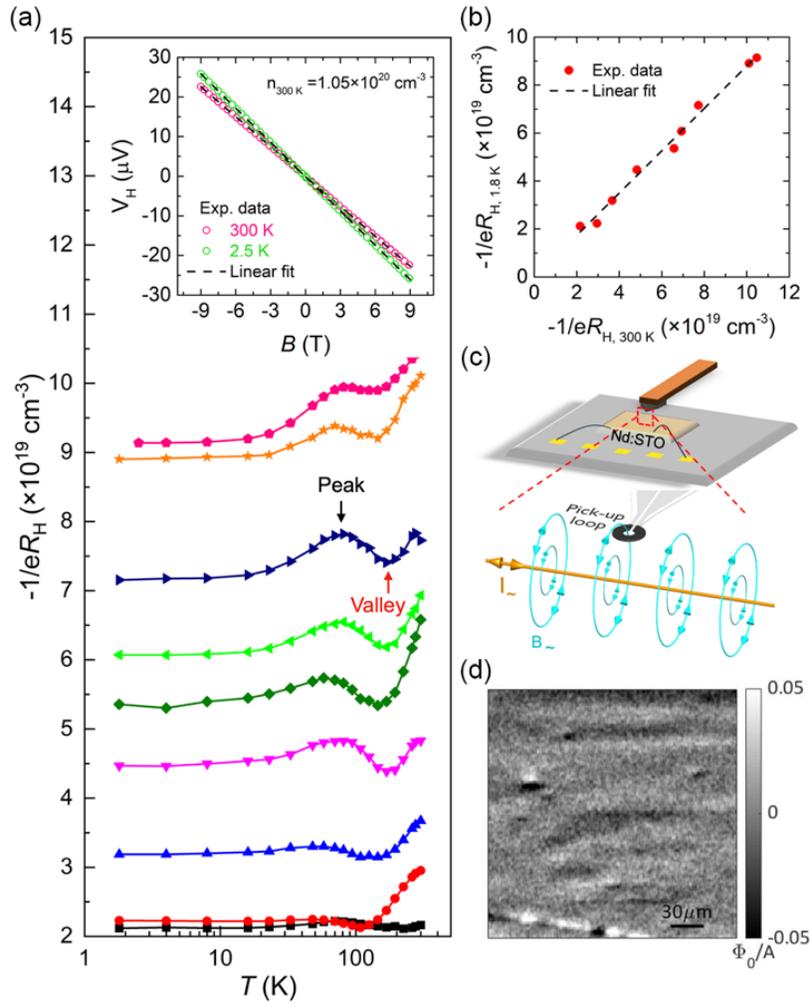

**Figure 3**: (a) The Hall carrier density $n_{Hall}$ ($-1/eR_H$) as a function of temperature for 60 nm Nd-doped STO/20 nm undoped STO/STO (001) samples for different doping densities. Inset shows linear Hall slopes at 2.5 K and 300 K for a representative sample with $n_{300\ K} = 1.05 \times 10^{20}$ cm$^{-3}$. (b) $n_{Hall}$ at 1.8 K as a function of $n_{Hall}$ at 300 K. A black dashed line shows a linear fit to the data yielding a slope of 0.88. (c) A schematic of the scanning SQUID measurement setup. (d) Scanning SQUID image of the sample with $n_{300\ K} = 4.8 \times 10^{19}$ cm$^{-3}$ showing the magnetic flux ($\phi_0$/A) generated by current flow at 4.5 K.

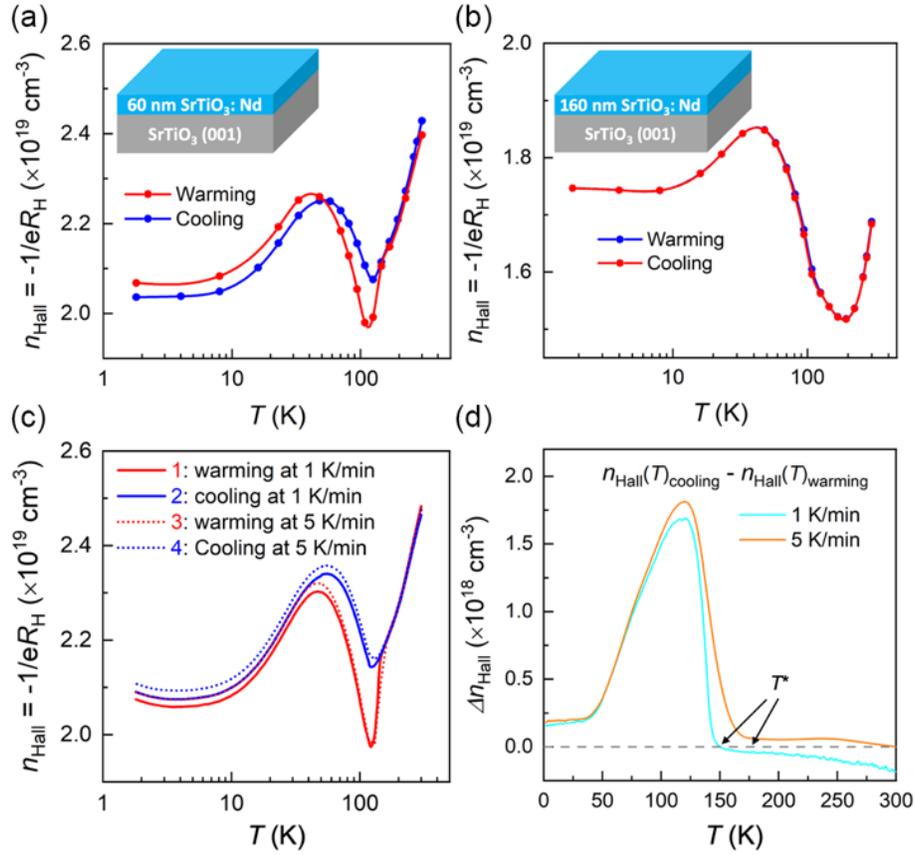

**Figure 4**: (a-b): Temperature dependence of the Hall carrier density $n_{Hall}$ of (a) 60 nm and (b) 160 nm Nd-doped STO sample with a similar $n_{Hall}$ during warming and cooling cycles. (c) $n_{Hall}$ and (d) difference of $n_{Hall}$ between cooling and warming cycles as a function of temperature of the same 60 nm Nd-doped STO sample measured at different warming/cooling rates. $T^*$ indicates the onset of difference in $n_{Hall}$ between warming and cooling cycles.

**References:**

1   K. Alex Müller, W. Berlinger, and E. Tosatti, *Z. Phys. B* **84** (2), 277 (1991).
2   E. K. H. Salje, O. Aktas, M. A. Carpenter, V. V. Laguta, and J. F. Scott, *Phys. Rev. Lett.* **111**, 247603 (2013).
3   J. Ruhman M. N. Gastiasoro, and R. M. Fernandes, Annals of Phys. **417**, 168107 (2020).
4   C. Collignon, X. Lin, C. W. Rischau, B. Fauqué, and K. Behnia, Annu. Rev. Condens. Matter Phys. (2018).
5   Ryan Russell, Noah Ratcliff, Kaveh Ahadi, Lianyang Dong, Susanne Stemmer, and John W. Harter, Phys. Rev. Materials **3** (9), 091401 (2019).
6   K. Ahadi, L. Galletti, Y. Li, S. Salmani-Rezaie, W. Wu, and S. Stemmer, Sci. Adv. **5** (4), eaaw0120 (2019).
7   S. Salmani-Rezaie, K. Ahadi, and S. Stemmer, Nano Lett. **20** (9), 6542 (2020).
8   Y. Tomioka, N. Shirakawa, K. Shibuya, and I. H. Inoue, Nat. Comm. **10** (1), 738 (2019).
9   C. W. Rischau, X. Lin, C. P. Grams, D. Finck, S. Harms, J. Engelmayer, T. Lorenz, Y. Gallais, B. Fauqué, J. Hemberger, and K. Behnia, Nat. Phys. **13** (7), 643 (2017).
10  X. Li, B. Fauque, Z. Zhu, and K. Behnia, Phys. Rev. Lett. **124** (10), 105901 (2020).
11  H. Unoki and T. Sakudo, J. Phys. Soc. Jpn. **23**, 546 (1967).
12  J. F. Scott, E. K. H. Salje, and M. A. Carpenter, Phys. Rev. Lett. **109** (18), 187601 (2012).
13  Y. Frenkel, N. Haham, Y. Shperber, C. Bell, Y. Xie, Z. Chen, Y. Hikita, H. Y. Hwang, E. K. H. Salje, and B. Kalisky, Nat. Mater. **16** (12), 1203 (2017).
14  B. Kalisky, E. M. Spanton, H. Noad, J. R. Kirtley, K. C. Nowack, C. Bell, H. K. Sato, M. Hosoda, Y. Xie, and Y. Hikita, Nat. mater. **12** (12), 1091 (2013).
15  J. A. Sulpizio M. Honig, J. Drori, A. Joshua, E. Zeldov & S. Ilani Nat. Mater. **12**, 1112 (2013).
16  H. J. Harsan Ma, S. Scharinger, S. W. Zeng, D. Kohlberger, M. Lange, A. Stöhr, X. Renshaw Wang, T. Venkatesan, R. Kleiner, J. F. Scott, J. M. D. Coey, D. Koelle, and Ariando, Phys. Rev. Lett. **116** (25), 257601 (2016).
17  E. Persky, N. Vardi, A. Mafalda R. V. L. Monteiro, T. C. van Thiel, H. Yoon, Y. Xie, B. Fauqué, A. D. Caviglia, H. Y. Hwang, K. Behnia, J. Ruhman, and B. Kalisky, Nat. Comm. **12**, 3311 (2021).
18  N. J. Goble, R. Akrobetu, H. Zaid, S. Sucharitakul, M.-H.Berger, A. Sehirlioglu, and X. P. A. Gao, Sci. Rep. **7**, 44361 (2017).
19  Y. Frenkel, N. Haham, Y. Shperber, C. Bell, Y. Xie, Z. Chen, Y. Hikita, H. Y. Hwang, and B. Kalisky, ACS Appl. Mater. Interfaces **8** (19), 12514 (2016).
20  Q. Tao, B. Loret, B. Xu, X. Yang, C. W. Rischau, X. Lin, B. Fauqué, M. J. Verstraete, and K. Behnia, Phys. Rev. B **94** (3), 035111 (2016).
21  H. Noad, P. Wittlich, J. Mannhart, and K. A. Moler, J. Supercond. Nov. Magn. **32** (4), 821 (2018).
22  H. Noad, E. M. Spanton, K. C. Nowack, H. Inoue, M. Kim, T. A. Merz, C. Bell, Y. Hikita, R. Xu, W. Liu, A. Vailionis, H. Y. Hwang, and K. A. Moler, Phys. Rev. B **94**, 174516 (2016).
23  R.A. Cowley, W. J. L. Buyers, and G. Dolling, Solid State Commun. **7** (1), 181 (1969).
24  G. Shirane and Y. Yamada, Phys. Rev. **177** (2), 858 (1969).
25  A. Verma, A. P. Kajdos, T. A. Cain, S. Stemmer, and D. Jena, Phys. Rev. Lett. **112** (21), 216601 (2014).

26 J. J. Zhou, O. Hellman, and M. Bernardi, Phys. Rev. Lett. **121** (22), 226603 (2018).
27 P. Wölfle and A. V. Balatsky, Phys. Rev. B **98**, 104505 (2018).
28 J. Ruhman and P. A. Lee, Phys. Rev. B **94**, 224515 (2016).
29 M. N. Gastiasoro, T. V. Trevisan, and R. M. Fernandes, Phys. Rev. B **101**, 174501 (2020).
30 D. Pelc S. Hameed, Z. W. Anderson, A. Klein, R. J. Spieker, L. Yue, B. Das, J. Ramberger, M. Lukas, Y. Liu, M. J. Krogstad, R. Osborn, Y. Li, C. Leighton, R. M. Fernandes, M. Greven, arXiv:2005.00514 [cond-mat.supr-con] (2020).
31 B. Jalan, R. E.- Herbert, N. J. Wright, and S. Stemmer, J. Vac. Sci. Technol. A **27** (3), 461 (2009).
32 B. Jalan, P. Moetakef, and S. Stemmer, Appl. Phys. Lett. **95** (3), 032906 (2009).
33 J. Son, P. Moetakef, B. Jalan, O. Bierwagen, N. J. Wright, R. E.- Herbert, and S. Stemmer, Nat. Mater. **9**, 482 (2010).
34 O. N. Tufte and P. W. Chapman, Phys. Rev. **155** (3), 796 (1967).
35 Y Kozuka, Y Hikita, C Bell, and HY Hwang, Applied Physics Letters **97** (1), 012107 (2010).
36 J. Yue Y. Ayino, T. Wang, B. Jalan, and V. Pribiag, J. Phys.: Condens. Matter **32**, 38LT (2020).
37 X. Lin, B. Fauqué, and K. Behnia, Science **349** (6251), 945 (2015).
38 A. Kumar, V. I. Yudson, and D. L. Maslov, Phys. Rev. Lett. **126**, 076601 (2021).
39 X. Lin, G. Bridoux, A. Gourgout, G. Seyfarth, S. Krämer, M. Nardone, B. Fauqué, and K. Behnia, Phys. Rev. Lett. **112** (20), 207002 (2014).
40 E. McCalla, M. N. Gastiasoro, G. Cassuto, R. M. Fernandes, and C. Leighton, Phys. Rev. Mater. **3** (2), 022001 (2019).
41 T. V. Trevisan, M. Schütt, and R. M. Fernandes, Phys. Rev. B **98** (9), 094514 (2018).
42 A. Joshua, S. Pecker, J. Ruhman, E. Altman, and S. Ilani, Nat. Comm. **3**, 1129 (2012).
43 Han Fu, K. V. Reich, and B. I. Shklovskii, Phys. Rev. B **94** (4) (2016).
44 Karthik Krishnaswamy, Burak Himmetoglu, Youngho Kang, Anderson Janotti, and Chris G. Van de Walle, Phys. Rev. B **95** (20) (2017).
45 P. Delugas, A. Filippetti, M. J. Verstraete, I. Pallecchi, D. Marré, and V. Fiorentini, Phys. Rev. B **88** (4), 045310 (2013).
46 D. V. Christensen, Y. Frenkel, P. Schütz, F. Trier, S. Wissberg, R. Claessen, B. Kalisky, A. Smith, Y. Z. Chen, and N. Prydsi, Phys. Rev. Appl. **9** (5), 054004 (2018).
47 L. G.-Ferreira, S. A. T. Redfern, E. Artacho, E. Salje, and W. T. Lee, Phys. Rev. B **81**, 024109 (2010).

# Anomalous Transport in High-Mobility Superconducting SrTiO$_3$ Thin Films

Jin Yue[1,a)], Yilikal Ayino[2], Tristan K. Truttmann[1], Maria N. Gastiasoro[2,3], Eylon Persky[4], Alex Khanukov[4], Dooyong Lee[1], Laxman R. Thoutam[1,5], Beena Kalisky[4], Rafael M. Fernandes[2], Vlad S. Pribiag[2], and Bharat Jalan[1,a)]

[1]Department of Chemical Engineering and Materials Science
University of Minnesota,
Minneapolis, Minnesota 55455, USA

[2]School of Physics and Astronomy
University of Minnesota,
Minneapolis, Minnesota 55455, USA

[3]Current affiliation: ISC-CNR and Department of Physics, Sapienza University of Rome,
Piazzale Aldo Moro 2, 00185, Rome, Italy

[4]Department of Physics and Institute of Nanotechnology and Advanced Materials
Bar-Ilan University,
Ramat Gan 5290002, Israel

[5]Current affiliation: Department of Electronics and Communications Engineering, SR University, Warangal Urban, Telangana, 506371, India

[a)] Corresponding author: yuexx129@umn.edu, bjalan@umn.edu

# I. Methods

**Film growth and characterization**

All samples were grown with a hybrid MBE approach. The details of the growth method is described elsewhere[1-3], but a brief description is provided here. The 5 mm × 5 mm SrTiO$_3$ (001) substrates (CrysTec GmbH, Germany) were heated to 900 °C (thermocouple temperature) in the hybrid MBE system (Scienta Omicron, Germany). Growth was preceded by 20 minutes of oxygen cleaning via 250 W RF oxygen plasma that achieves a background oxygen pressure of 5 × 10$^{-6}$ Torr (Mantis, UK). Strontium and neodymium were provided via thermal sublimation from an effusion cell. Strontium was supplied at 472 °C to achieve a beam equivalent pressure (BEP) of 4 × 10$^{-8}$ Torr, whereas the neodymium effusion cell temperature was varied between 780 and 980 °C to control the dopant concentration. During growth, oxygen was supplied using the same oxygen plasma parameters that were used for oxygen cleaning.

High-Energy Electron Diffraction (RHEED) was used to characterize the sample growth *in-situ*, atomic force microscopy (AFM) was used to characterize the samples surface *ex-situ*. High-resolution X-ray diffraction (HR-XRD) data were collected with a PANalytical X'Pert Pro thin film diffractometer with a Cu parabolic mirror and germanium 4-bounce monochromator. All transport data were collected in the van der Pauw geometry in the temperature- and magnetic field-controlled environment provided by a DynaCool physical property measurements system, (Quantum Design, USA). Magnetic field was swept between ± 9 T.

A buffer layer of insulating 20 nm STO was grown on each substrate prior to growing doped layer to minimize substrate surface effect. Films grown without doping were insulating with no measurable conductivity indicating no contributions of oxygen vacancies to the electrical transport. For scanning SQUID, an AC current was applied through the adjacent corners in the vdP geometry. The generated magnetic flux was measured using a 0.75 μm sensing loop (pick-up loop) via lock-in mechanism. By mapping the flux near the surface of the sample, a 2D map describing the local current flow distribution was generated.

**Residual resistivity vs. carrier density calculation across Lifshitz transition**

We consider a toy two-band model with 3D isotropic dispersions illustrated in figure 2c. For chemical potential $\mu < \mu_c$, only the bottom band is occupied. At $\mu = \mu_c$ the second band crosses the Fermi surface. We calculate the dc conductivity due to elastic scattering by impurities at zero temperature and study its behavior when the Lifshitz transition takes place. The Kubo formula for conductivity in multi-band system is

$$\sigma = \frac{e^2}{3} \sum_\alpha v_{F,\alpha}^2 N_\alpha \tau_\alpha.$$

Here $v_{F,\alpha}^2$ and $N_\alpha$ are the velocity and density of states at the Fermi level of band $\alpha$ ($\alpha$ = 1,2). For isotropic intra-band and inter-band scattering the transport relaxation time can be approximated by the lifetime, which in the first Born approximation is given by $\tau_\alpha^{-1} = 2\pi n_{imp} V_{intra}^2 (N_\alpha + \gamma^2 N_\beta)$ with $\beta \neq \alpha$ for a dilute density of impurities $n_{imp}$. Finally, $V_{inter}$ and $V_{intra}$ describes the inter-band and intra-band scattering potentials, respectively, and we have introduced the ratio $\gamma = \frac{V_{inter}}{V_{intra}}$. The final expression for the conductivity is then

$$\sigma = A \frac{\mu}{n_{imp}} \left[ \theta(-\bar{\mu} + 1) + \theta(\bar{\mu} - 1) \left( \frac{1}{1 + \gamma^2 \sqrt{1 - \bar{\mu}^{-1}}} + \frac{1 - \bar{\mu}^{-1}}{1 + \frac{\gamma^2}{\sqrt{1 - \bar{\mu}^{-1}}}} \right) \right],$$

where $A = \frac{e^2}{3\pi V_{intra}^2 m}$ and we have introduced the normalized chemical potential $\bar{\mu} = \frac{\mu}{\mu_c}$. The Heaviside functions $\theta(\pm\bar{\mu} \mp 1)$ separate the single-band transport regime from the two-band regime. We plot in figure 2d the renormalized resistivity $\rho/\rho_c$ ($\rho_c = \frac{3\pi V_{intra}^2 n_{imp} m}{e^2 \mu_c}$) as a function of the carrier density $n = \frac{(2m\mu_c)^{3/2}}{3\pi^2}[\bar{\mu}^{3/2} + (\bar{\mu}-1)^{3/2}\theta(\bar{\mu}-1)]$, in which the chemical potential $\mu = \mu_c$ corresponds to the critical density $n = n_c$ where the Lifshitz transition takes place. In order to capture the weak dependence of $\rho$ on doping above (and below) the Lifshitz transition, we modeled the density of impurities to follow $n_{imp} = n^{2/3}$. If we had instead kept $n_{imp}$ doping independent, or assumed it to depend linearly on $n$, the residual resistivity would still increase immediately across the Lifshitz transition for $\gamma = 0.5, 1$.

The main conclusion is that if the inter-band scattering is much weaker than the intra-band scattering, the resistivity drops when the Lifshitz transition takes place. On the other hand, in the $V_{intra} = V_{inter}$ case one should expect a sharp increase of the resistivity. The experimental data agrees with a scenario where intra-band scattering dominates, $V_{inter} \ll V_{intra}$.

## II. Temperature-dependence of pre-factor $A$ and Fermi temperature

Figure S1a follows a $A \sim n^{-1}$ relationship, consistent with previous literature reports[4-7]. This inverse relationship between the pre-factor $A$ and the carrier density $n$ was argued to be an indication of a carrier density independent scattering rate[4,8]. For $T \leq 100$ K, a linear $\rho$ vs $T^2$ behavior was observed in all the samples, which is commonly attributed to standard Fermi liquids (FL) behavior. However, given the suppression of Umklapp processes due to the small sizes of the Fermi surfaces, the origin of this quadratic temperature dependence in doped STO remains widely debated[4,5,8-12].

Indeed, in our samples the Fermi temperature $T_F$ (estimation of $T_F$ is shown in Fig. S1b) is roughly the same as the measurement temperature, placing the system in a regime where the mechanism for $T^2$ behavior of the resistivity predicted by standard Fermi liquid is not applicable.

Other inconsistencies with standard FL transport theory have also been discussed in the literature[4,5,13], and future study should be carried out for better understanding. We note, however, that as discussed in ref.[11], specific heat measurements in single crystals show the existence of well-defined quasi-particles with a weak mass renormalization.

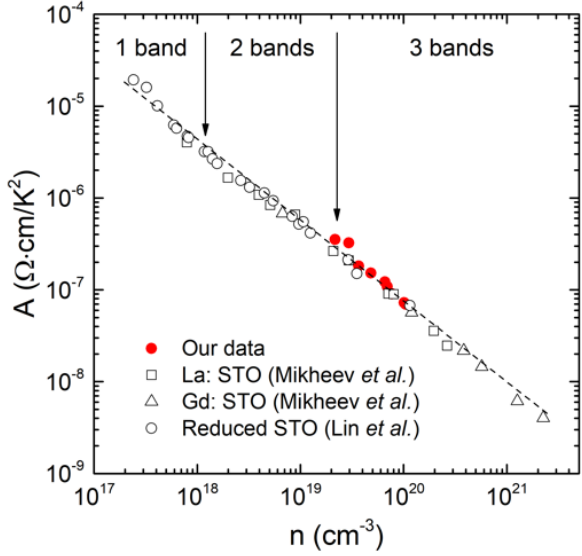 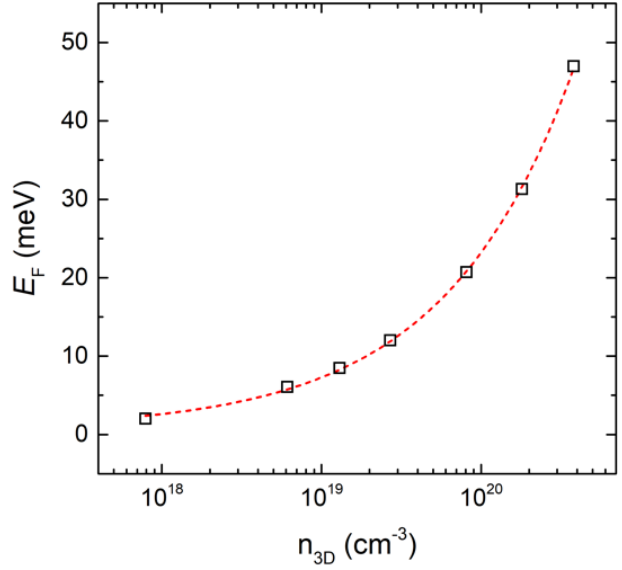

**Figure S1a**: A combined plot of the pre-factor $A$ vs carrier density $n$ for our data as well as the data from the literature[4,5]. The pre-factor $A$ monotonically decreases with the carrier density $n$ and follows $A \sim n^{-1}$ relationship over about 4 orders of magnitudes.

**Figure S1b**: Calculated Fermi Energy $E_F$ vs. 3D carrier density $n_{3D}$ of $SrTiO_3$ based on the DFT-calculated band structure of the tetragonal phase using the tight-binding model. First, we calculated the Fermi energies for a few carrier densities (open black squares) within the tetragonal tight-binding model by self-consistent calculation of a chemical potential corresponding to each carrier density. Then we interpolate these calculated data points with a polynomial function (red short dash line) to get the $E_F$ vs. $n_{3D}$ dependence. A factor of two mass enhancement has been included by hand into this non-interacting model. This factor was determined by comparing the Sommerfeld coefficient from specific heat measurements and the calculated one from this model[11].

## III. Single-band transport and the presence of anomalous behavior around AFD transition

Figure S2 shows the experimentally measured Hall carrier density, $n_{Hall}$ as a function of temperature in a 1280 nm thick La-doped STO film and 1060 nm Nd-doped STO film. Despite low carrier density where a single band is occupied in STO, both samples showed a similar anomalous behavior around 100 K suggesting that its origin is not due to a possible nonlinear Hall effect owing to the multiband conduction.

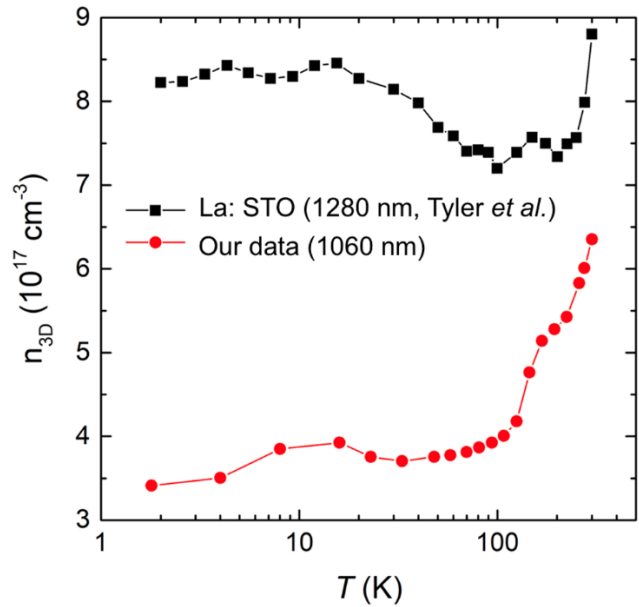

**Figure S2**: Carrier density vs. temperature plot for a thick Nd-doped STO and La-doped STO sample (data from literature[14]), which shows qualitatively the same evolution trend as described in the main text indicating that multiband effect is not responsible for this universal carrier density evolution trend.

### IV. Temperature Dependent Hall Scattering Factor in SrTiO$_3$

Figure S3 shows the temperature dependence of the inverse Hall scattering factor ($r_H$) from first-principles calculations reported in the literature[15]. These calculated values were interpolated with a polynomial function to convert the experimental $n_{Hall}$ to the true 3D carrier density ($n_{3D}$). Since the dependence of $r_H$ is weak on the carrier density, we used a polynomial function from $2 \times 10^{19}$ cm$^{-3}$ to extract $n_{3D}$ for all the samples in this work.

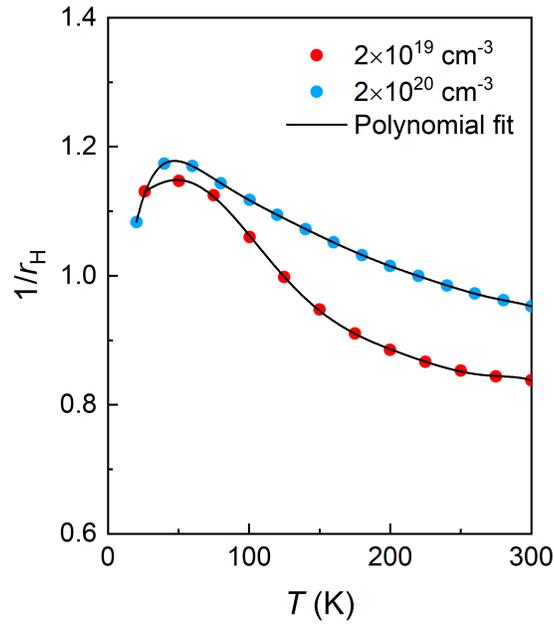

**Figure S3**: The temperature dependence of $1/r_H$ of STO when $n_{3D}$ is $2\times10^{19}$ and $2\times10^{20}$ cm$^{-3}$.[15] The black lines are polynomial fits to the data.

Figure S4 shows $n_{Hall}$ ($T$), the extracted $n_{3D}$ ($T$) and an Arrhenius plot of $n_{3D}$. The anomalous behavior (peak-and-valley pattern) observed in $n_{Hall}$ ($T$) does not appear in the $n_{3D}$ vs $T$ curves after correcting for the Hall scattering factor. These results suggest that this anomalous behavior is likely related to the temperature dependence of the Hall scattering factor. However, $n_{3D}$ ($T$) showed a monotonic decrease with decreasing temperature. Although these are high carrier densities, one may still argue the possibility of the thermally activated transport. To show that it is not the case, we plotted $n_{3D}$ ($T$) on an Arrhenius plot (figure S4c) to show that the drop in $n_{3D}$ with decreasing temperature cannot be explained by a thermally activated behavior. Moreover, around 100 K in figure S4b, $n_{3D}$ shows a dramatic change of slope suggesting that this behavior may be related to the AFD transition in STO.

Recent studies showed that the domain walls formed in this phase transition can be polar or even ferroelectric[16-19], and these polar nanoregions are argued to be responsible for the ferroelectricity observed in STO (even in the strain-free case!)[20]. It is conceivable that apart from the changes in phonon related scatterings, the AFD transition can also impact the electronic transport properties of STO via the change in local carrier density and screening.

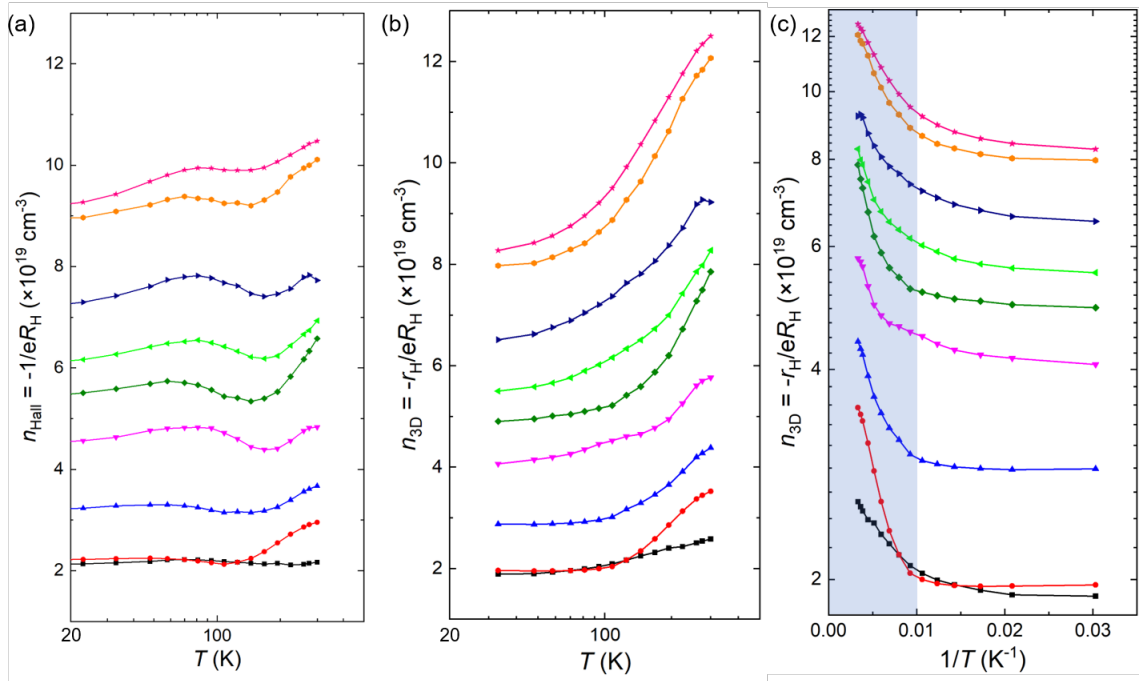

**Figure S4**: (a) The nominal Hall carrier density $n_{Hall}$ ($-1/eR_H$) and (b) 3D carrier density $n_{3D}$ ($-r_H/eR_H$) as a function of temperature for the aforementioned Nd-doped STO sample series with 60 nm active layer thickness. (c) $n_{3D}$ is plotted against $1/T$ on a semi-log scale to check for Arrhenius behavior.

# V. References

1. B. Jalan, R. E.- Herbert, N. J. Wright, and S. Stemmer, J. Vac. Sci. Technol. A **27** (3), 461 (2009).
2. B. Jalan, P. Moetakef, and S. Stemmer, Appl. Phys. Lett. **95** (3), 032906 (2009).
3. J. Son, P. Moetakef, B. Jalan, O. Bierwagen, N. J. Wright, R. E.- Herbert, and S. Stemmer, Nat. Mater. **9**, 482 (2010).
4. E. Mikheev, S. Raghavan, J. Y. Zhang, P. B. Marshall, A. P. Kajdos, L. Balents, and S. Stemmer, Sci. Rep. **6**, 20865 (2016).
5. X. Lin, B. Fauqué, and K. Behnia, Science **349** (6251), 945 (2015).
6. D. van der Marel, J. L. M. vanMechelen, and I. I. Mazin, Phys. Rev. B **84** (20), 205111 (2011).
7. Y. Tomioka, N. Shirakawa, K. Shibuya, and I. H. Inoue, Nat. Comm. **10** (1), 738 (2019).
8. S. Stemmer and S. J. Allen, Rep. Prog. Phys. **81** (6), 062502 (2018).
9. C. Collignon, X. Lin, C. W. Rischau, B. Fauqué, and K. Behnia, Annu. Rev. Condens. Matter Phys. (2018).
10. X. Lin, C. W. Rischau, L. Buchauer, A. Jaoui, B. Fauqué, and K. Behnia, npj Quantum Materials **2** (1), 41 (2017).
11. E. McCalla, M. N. Gastiasoro, G. Cassuto, R. M. Fernandes, and C. Leighton, Phys. Rev. Mater. **3** (2), 022001 (2019).
12. A. Kumar, V. I. Yudson, and D. L. Maslov, Phys. Rev. Lett. **126**, 076601 (2021).
13. M. W. Swift and C. G. Van de Walle, Eur. Phys. J. B **90** (8), 151 (2017).
14. T. A. Cain, A. P. Kajdos, and S. Stemmer, Appl. Phys. Lett. **102** (18), 182101 (2013).
15. P. Delugas, A. Filippetti, M. J. Verstraete, I. Pallecchi, D. Marré, and V. Fiorentini, Phys. Rev. B **88** (4), 045310 (2013).
16. J. F. Scott, E. K. H. Salje, and M. A. Carpenter, Phys. Rev. Lett. **109** (18), 187601 (2012).
17. E. K. H. Salje, O. Aktas, M. A. Carpenter, V. V. Laguta, and J. F. Scott, *Phys. Rev. Lett.* **111**, 247603 (2013).
18. Y. Frenkel, N. Haham, Y. Shperber, C. Bell, Y. Xie, Z. Chen, Y. Hikita, H. Y. Hwang, E. K. H. Salje, and B. Kalisky, Nat. Mater. **16** (12), 1203 (2017).
19. H. J. Harsan Ma, S. Scharinger, S. W. Zeng, D. Kohlberger, M. Lange, A. Stöhr, X. Renshaw Wang, T. Venkatesan, R. Kleiner, J. F. Scott, J. M. D. Coey, D. Koelle, and Ariando, Phys. Rev. Lett. **116** (25), 257601 (2016).
20. H. W. Jang, A. Kumar, S. Denev, M.D. Biegalski, P. Maksymovych, C.W. Bark, C.T. Nelson, C.M. Folkman, S. H. Baek, N. Balke, C.M. Brooks, D. A. Tenne, D. G. Schlom, L. Q. Chen, X. Q. Pan, S.V. Kalinin, V. Gopalan, and C. B. Eom, Phys. Rev. Lett. **104** (19), 197601 (2010).